# Agent-Based Modeling and Simulation of Connected and Automated Vehicles Using Game Engine: A Cooperative On-Ramp Merging Study

**Ziran Wang (Corresponding Author)**
Department of Mechanical Engineering, University of California, Riverside
1084 Columbia Ave, Riverside, CA 92507, USA
Phone: (626)271-3096; Fax: (951)781-5790; Email: zwang050@ucr.edu

**BaekGyu Kim, Ph.D.**
Toyota InfoTechnology Center, U.S.A. Inc.
465 N Bernardo Ave, Mountain View, CA 94043, USA
Phone: (650)694-4100; Fax: (650)694-4901; Email: bkim@us.toyota-itc.com

**Hiromitsu Kobayashi**
Toyota InfoTechnology Center, U.S.A. Inc.
465 N Bernardo Ave, Mountain View, CA 94043, USA
Phone: (650)694-4100; Fax: (650)694-4901; Email: hkobayashi@us.toyota-itc.com

**Guoyuan Wu, Ph.D.**
Center for Environmental Research and Technology, University of California, Riverside
1084 Columbia Ave, Riverside, CA 92507, USA
Phone: (951)781-5630; Fax: (951)781-5790; Email: gywu@cert.ucr.edu

**Matthew J. Barth, Ph.D.**
Department of Electrical and Computer Engineering, University of California, Riverside
1084 Columbia Ave, Riverside, CA 92507, USA
Phone: (951)781-5782; Fax: (951)781-5790; E-mail: barth@ece.ucr.edu

**ABSTRACT**
Agent-based modeling and simulation (ABMS) has been a popular approach to modeling autonomous and interacting agents in a multi-agent system. Specifically, ABMS can be applied to connected and automated vehicles (CAVs), since CAVs can be driven autonomously with the help of on-board sensors, and cooperate with each other through vehicle-to-everything (V2X) communications. In this work, we apply ABMS to CAVs using the game engine Unity3D, taking advantage of its visualization capability and other capabilities. Agent-based models of CAVs are built in the Unity3D environment, where vehicles are enabled with connectivity and autonomy by C#-based scripting API. We also build a simulation network in Unity3D based on the city of Mountain View, California. A case study of cooperative on-ramp merging has been carried out with the proposed distributed consensus-based protocol, and then compared with the human-in-the-loop simulation where the on-ramp vehicle is driven by four different human drivers on a driving simulator. The benefits of introducing the proposed protocol are evaluated in terms of travel time, energy consumption, and pollutant emissions. It is shown from the results that the proposed cooperative on-ramp merging protocol can reduce average travel time by 7%, reduce energy consumption and pollutant emissions by 8% and 58%, respectively, and guarantee the driving safety when compared to the human-in-the-loop scenario.

**INTRODUCTION**
Agent-based modeling and simulation (ABMS) focuses on microscale models that simulate the simultaneous operations and interactions of multiple agents (*1*). There is no universal definition of the term "agent", however, some certain characteristics are often shared by agents from a practical modeling standpoint (*2*). Those characteristics include: a) Identifiable, with rules governing their decision-making capabilities; b) Interactive, with the ability to recognize and distinguish the traits of other agents; c) Goal-directed, with goals to achieve with respect to their behaviors; d) Autonomous, with the capability to function independently in their environment; e) Flexible, with the ability to learn and adapt their behaviors over time based on experience. Given the fact that connected and automated vehicles (CAVs) can fulfill above characteristics to some extent, ABMS is considered an attractive approach to evaluating transportation systems comprised of CAVs.

Many different tools for the modeling and simulation of CAVs are available. From the traditional four-step models, to state-of-the-art agent-based models, they all have their unique advantages and thus suited for different research purposes (*3*). The emergence of the CAV technology triggers difficult system modeling problems. Traditional development tools that consider only one target vehicle can no longer be adopted, since we need to also consider a CAV's surrounding environment, such as other vehicles, road, infrastructure, and pedestrians. Generally, game engines provide complex modeling of virtual reality environments, and also allow users (i.e., game players) to dynamically involve in the game during the simulation process. Therefore, in this work, we adopt the Unity3D game engine to conduct ABMS of CAVs in a case study of cooperative on-ramp merging.

The structure of this paper is as follows: The following section reviews some game engine-based modeling and simulation work of CAVs, as well as different related work about on-ramp merging systems. Section 3 demonstrates the proposed methodology of distributed consensus-based cooperative on-ramp merging, including the vehicle sequencing protocol and vehicle longitudinal control protocol. The modeling of CAVs in the proposed cooperative on-ramp merging case and the simulation results are included in Section 4. The last section concludes the paper, describing some potential next steps.

**BACKGROUND AND RELATED WORK**
**Game Engine-Based Modeling and Simulation of Vehicles**
Traditionally, game engines are software systems designed for the creation and development of video games. A typical game engine consists of different core components, including a rendering engine (often called "renderer") for two-dimensional or three-dimensional graphics, a physical engine for collision detection and its response, and a scene graph for the management of models, scripting, sound, threading, and networking, among other things. Nowadays, there are plenty of game engines available for product development and testing, which can be classified into two types: a) Commercial game engines such as Unity3D, Unreal, and Blender, which are powerful but also relatively expensive; and b) Open-source game engines such as Quake 3, Delta3D, and Nebula, where the performances of this type of engines usually depend on particular applications.

In this work, we conduct ABMS of CAVs using the Unity3D game engine, which integrates a custom rendering engine with the Nvidia PhysX physics engine and Mono, the open source implementation of Microsoft's .NET libraries (*4*). Unity 3D has been widely used to build simulation platforms. Graighead *et al.* implemented the Search and Rescue Game Environment (SARGE) with Unity3D. Robotic vehicles in this environment are implemented with various sensors, such as 3D camera, GPS, odometry, inertial measurement (IMU), and planar laser ranging (*5*). KTH Royal Institute of Technology in Sweden conducted several research work on the visualization of truck platooning using Unity3D (*6*, *7*). Toyota

InfoTechnology Center in USA also contributed a series of work to the vehicle prototyping research by Unity3D. Yamaura *et al.* built a virtual prototype of Advanced Driver Assistance System (ADAS) with a closed-loop simulation framework that consists of four tools: Unity3D, Simulink OpenMETA, and Dymola (*8*). Kim *et al.* proposed several associated directions and potential approaches of testing autonomous vehicle software in virtual prototyping environment using Unity3D, from the perspective of test criteria and test case generation (*9*). As an extended work of that, Dai *et al.* presented a co-simulation tool-chain for the automated optimization of various parameters in the virtual prototyping environment (*10*). More detailed introductions of Unity3D are included in Section 4.

Some of the reasons that we choose Unity3D rather than other game engines and simulation tools in this work are listed below:
a) Graphics and visualization: Since Unity3D is designed for developers to develop 3D video games, it has an impressive capability of graphics and visualization. It streamlines the demonstration of the proposed CAV technology to the audience, especially to the general public (without knowing technical details). This is the primary reason why we selected Unity3D for performing ABMS of CAVs.
b) Integration of driving simulator: Unity3D provides easy access to changing the input equipment, which makes it possible to integrate our driving simulator hardware. Since we want to compare the proposed CAV technology with the baseline, a driving simulator hardware with human-in-the-loop simulation is more realistic than simply applying some human driver models in the simulation.
c) Asset store: Unity3D has an official asset store where Unity3D developers and users can upload and download different Unity3D assets, which allows Unity3D users to develop their own game environment based on others' existing work, instead of building things from scratch.
d) Documentation and community: Unity3D provides thorough, well-organized and easy-to-read documentation covers how to use each component in Unity3D, and an online commUnity3D website for all Unity3D users to ask and answer questions.

**On-Ramp Merging**
On-ramp merging, especially for highway, has been an interesting topic with extensive studies by researchers around the world. A literature review on the coordination of CAVs merging at highway on-ramps was accomplished by Rios-Torres *et al.*, which summarizes the developments and research trends in this research field (*11*). It can be noted that optimal control approach has been adopted by many of the recent on-ramp merging work. Rios-Torres *et al.* presented an optimization framework and an analytical close-form solution to allow online coordination of merging vehicles (*12*). A proactive optimal merging strategy was proposed by Awal *et al.* to compute the optimal merging order for vehicles coming from mainline and on-ramp, which introduces different benefits in terms of energy consumption, merging efficiency, and traffic flow (*13*). Raravi *et al.* came up with an approach to optimize the time-to-conflict-zone for every vehicle once their merging sequences are defined (*14*). Model Predictive Control (MPC) scheme was adopted by Cao *et al.* to generate a cooperative merging path for vehicles to merge smoothly on the mainline (*15*).

Besides the optimization-based methods, some other approaches have also been used for on-ramp merging. Milanes *et al.* developed fuzzy-logic method to allow vehicles to merge from the ramp to mainline fluidly without causing congestion on the ramp, while changing the speed of mainline vehicles to minimize the effect on the already-congested mainline (*16*). Marinescu *et al.* developed a slot-based algorithm for merging vehicles to cooperate with each other in a highly efficient manner (*17*). Uno *et al.* used the virtual vehicle platooning concept to map a virtual vehicle onto the mainline before it actually merges (*18*). Lu *et al.* adopted a centralized controller to interchange relevant information with merging vehicles, and each merging vehicle conducts its own control actions to achieve the assigned time and reference speed requirements (*19*, *20*).

Limitations of aforementioned on-ramp merging methods include: 1. Some of the methods are not always suitable for real-time implementation due to the difficulty to find a solution; 2. A vehicle is considered in a single form, making it difficult to extend to a vehicle string; 3. Benefits of energy efficiency and pollutant emissions are not shown. In next section, the cooperative on-ramp merging system we introduce addresses these issues.

## COOPERATIVE ON-RAMP MERGING
In this section, we introduce the distributed consensus-based cooperative on-ramp merging system that will be modeled and simulated in the Unity3D game engine. This system is partially built on top of our previous work (*21*). This section consists of three major parts: System architecture and specifications, vehicle sequencing protocol, and distributed consensus-based longitudinal control protocol.

### System Architecture and Specifications
The proposed distributed consensus-based cooperative on-ramp merging system utilizes V2X communications of CAVs, which means vehicles can communicate with each other through V2V communications, and with the infrastructure through V2I communications. The illustrative system architecture is shown in FIGURE 1. When merging vehicles from the upstream of on-ramp and highway enter the V2I communication range of the infrastructure, they send the infrastructure their own information measured by on-board sensors. That information includes but is not limited to current acceleration, speed, and global position of the vehicle. Then, the computer connected to the infrastructure processes all the information gathered from on-coming vehicles within a certain time interval, and assigns a series of sequence identification numbers to different vehicles based on the vehicle sequencing protocol. Vehicles will retrieve those sequence identification numbers from the infrastructure at the next time step. Once a vehicle gets its own sequence identification number, it can automatically match its predecessor (which is not necessarily the one in front), and cooperate with that vehicle based on the proposed longitudinal control protocol. It shall be noted that, all vehicles involved in the proposed system are CAVs, and only the longitudinal control is analyzed.

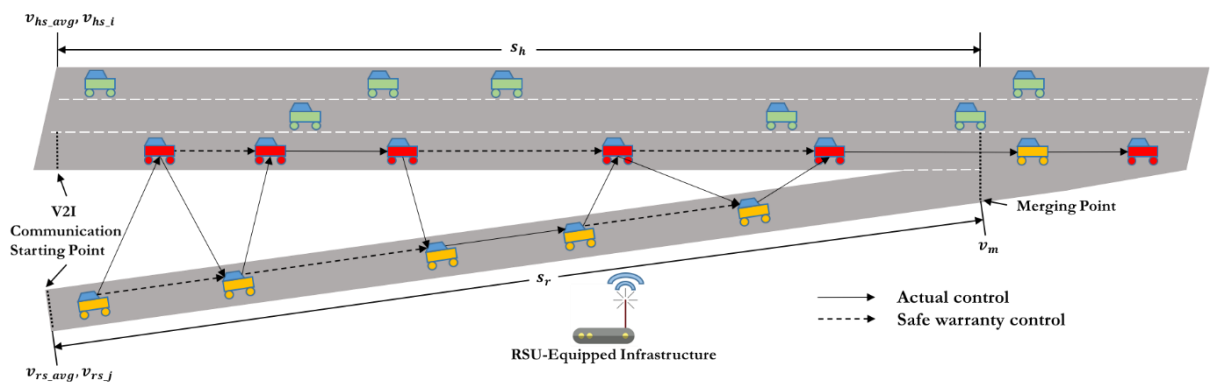

**FIGURE 1 Illustrative architecture of the cooperative on-ramp merging system.**

Parameters used in this work are defined in TABLE 1.

**TABLE 1 Annotation of Parameters in the Cooperative On-Ramp Merging System**

| Parameter | Annotation |
|---|---|
| $v_{lim}$ | Highway speed limit |
| $v_{hs\_i}$ | Speed of vehicle $i$ when reaching the V2I communication starting point on the rightmost lane of highway |
| $v_{hs\_avg}$ | Average speed of vehicles when reaching the V2I communication starting point on the rightmost lane of highway during a past time window |
| $v_{rs\_j}$ | Speed of vehicle $j$ when reaching the V2I communication starting point on the on-ramp |
| $v_{rs\_avg}$ | Average speed of vehicles when reaching the V2I communication starting point on the on-ramp during a past time window |
| $a_{max}$ | Maximum acceleration of vehicles without compromising safety and comfort |
| $s_h$ | Distance to the merging point when the highway vehicle sent its information to the infrastructure |
| $s_r$ | Distance to the merging point when the ramp vehicle sent its information to the infrastructure |
| $s_{acc}$ | Distance travelled by on-ramp vehicles accelerating from $v_{rs\_avg}$ to $v_{lim}$ |
| $v_{rm\_max}$ | Maximum reachable speed of on-ramp vehicles when reaching the merging point |
| $t_{h\_i}$ | Estimated time to arrive at the merging point by highway vehicle $i$ |
| $t_{r\_j}$ | Estimated time to arrive at the merging point by on-ramp vehicle $j$ |
| $t_{head\_safe}$ | Estimated time spent by vehicle $j$ travelling the distance of $s_r$ |
| $t_{head\_V2V}$ | V2V connection time headway |
| $v_m$ | Estimated merging speed |

**Vehicle Sequencing Protocol**

The objective of vehicle sequencing protocol is to arrange upstream merging vehicles with a preset sequence, so they can cooperate with their corresponding predecessors before coming to the merging point. Upon retrieving information from coming CAVs, the infrastructure firstly calculates different vehicles' estimated arrival time at the merging point, then sorts those time with a certain frequency. It shall be noted that the estimated arrival time is only a reference of the vehicle sequence and oftentimes that is difficult to be precisely calculated beforehand, therefore, so we assume simplified linear acceleration profile of vehicles. Also, the sorting frequency is relatively low (e.g., 0.2 Hz) compared to the system updating frequency (e.g., 10 Hz), since frequently changing vehicles' sequence introduces instability of vehicles' longitudinal control.

Since vehicles coming from the upstream of the on-ramp are most likely to accelerate from a relatively low speed to a relatively high speed, we want to calculate the estimated maximum reachable speed of on-ramp vehicles. Given the highway speed limit $v_{lim}$, the distance of on-ramp vehicles accelerating to highway speed limit can be calculated as

$$S_{acc} = \frac{(v_{rs-avg} + v_{lim})}{2} \cdot \frac{v_{lim} - v_{rs-avg}}{a_{max}} = \frac{v_{lim}^2 - v_{rs-avg}^2}{2a_{max}} \quad (1)$$

If $S_{acc} \geq S_r$, which means on-ramp vehicles can accelerate to highway speed limit $v_{lim}$ before merging, then the maximum reachable speed $v_{rm\_max}$ will simply be the highway speed limit. If $S_{acc} < S_r$, then the maximum reachable speed will be calculated as

$$v_{rm\_max} = v_{rs\_avg} + a_{max} \cdot \frac{-v_{rs\_avg} + \sqrt{v_{rs\_avg}^2 + 2a_{max}s_r}}{a_{max}} = \sqrt{v_{rs\_avg}^2 + 2a_{max}s_r} \quad (2)$$

We then compare $v_{rm\_max}$ with $v_{hs\_avg}$, and set the estimated merging speed $v_m$ to be the minimum of these two values. Specifically, the estimated time to arrive at the merging point of highway vehicle $i$ and on-ramp vehicle $j$ can be calculated as

$$\begin{cases} t_{h\_i} = \frac{s_h}{v_{hs\_i}} \\ t_{r\_j} = \frac{2a_{max}s_r + (v_{hs\_avg} - v_{rs\_j})^2}{2a_{max}v_{hs\_avg}} \end{cases}, when\ v_m = v_{hs\_avg} \quad (3)$$

$$\begin{cases} t_{h\_i} = \dfrac{2a_{max}(s_h-s_r)-(v_{hs\_i}^2+v_{rs_{avg}}^2)+2v_{hs\_i}\sqrt{v_{rs\_avg}^2+2a_{max}s_r}}{2a_{max}\sqrt{v_{rs\_avg}^2+2a_{max}s_r}} \\ \\ t_{r\_j} = \dfrac{-v_{rs\_j}+\sqrt{v_{rs\_j}^2+2a_{max}s_r}}{a_{max}} \end{cases}, when\ v_m = v_{rm\_max} \quad (4)$$

Once the estimated time to arrive at the merging point is calculated, it can be added up to the current time and then gets the estimated arrival time. Different from the estimated time to arrive, which is a time interval (e.g., 18.0 s), the estimated arrival time is the time of day (e.g., 20:00:18, where we assume current time is 20:00:00).

There is a possibility that the estimated arrival time of a following vehicle is not later than a preceding vehicle on the same lane. In that case, the estimated arrival time of this following vehicle should be $t_{head\_safe}$ later than preceding vehicle's value. It is also possible that a highway vehicle and an on-ramp vehicle have the same value of estimated arrival time, then we can set the estimated arrival time of the on-ramp vehicle to be $t_{head\_safe}$ later than the preceding vehicle's value. Upon the estimated arrival time of all vehicles, which comes into the V2I communication range during this current sorting interval (e.g., 5 s), are finalized, all the values can be sorted from earliest to latest, and be associated with different sequence identification numbers. Each vehicle will retrieve its sequence identification number from the infrastructure through V2I communications at the next time step, and identifies its predecessor based on that information.

**Distributed Consensus-Based Longitudinal Control Protocol**
Once each vehicle finds its predecessor, the proposed distributed consensus protocol can be applied to control the longitudinal movement. Here we use the idea of Cooperative Adaptive Cruise Control (CACC) to control the vehicle (*22*). If a vehicle and its predecessor (determined by their sequence identification number) are in the same lane, then this is exactly the case of predecessor-follower CACC, where we adopt our previous proposed distributed consensus-based CACC algorithm for this vehicle to follow its predecessor (*23, 24*)

$$a_k = -\delta[(s_k - s_p + s_{head}) + \gamma(v_k - v_p)] \quad (5)$$

where $a_k$ is the acceleration of vehicle $k$; $\delta$ and $\gamma$ are tuning coefficients; $s_k$ is the relevant position of vehicle $k$ to the merging point; $s_p$ is the relevant position of vehicle $p$ to the merging point; $v_k$ is the longitudinal speed of vehicle $k$; $v_p$ is the longitudinal speed of vehicle $p$; $s_{head} = \min(v_k t_{head_{safe}}, s_{head\_safe})$ is the desired distance headway, where $s_{head\_safe}$ is the safety distance headway.

If a vehicle and its predecessor (determined by their sequence identification number) are on different lanes, then the predecessor can be considered as a "ghost" predecessor and projected on the same lane of the vehicle. The "ghost" vehicle shares all the same parameters (except for the lateral position) with the actual predecessor. The ego vehicle can still use equation (5) to follow this "ghost" predecessor, but meanwhile also checks the time-to-collision value with respect to its physical preceding vehicle on the same lane by

$$t_{collision} = \frac{v_k - v_{p\_physical}}{s_{gap}} \quad (6)$$

where $v_{p\_physical}$ is the longitudinal speed of the physical preceding vehicle, and $s_{gap}$ is the gap distance between the vehicle and its physical preceding vehicle. Both values can be measured by vehicle's on-board sensors such as radar. Whenever the time-to-collision value hits a set lower bound (e.g., 2 s), the distributed consensus-based control will be deactivated, and the internal car-following controller takes over control of the vehicle. This time-to-collision checking process is a fail-safe logic to prevent rear-end collision caused by impacted V2V communications.

## AGENT-BASED MODELING AND SIMULATION
### Agent-Based CAV Modeling

A vehicle is modeled as a game object in Unity3D, which always includes a rigid body and its associated colliders. By applying forces or torque to the vehicle's rigid body, the vehicle will start to move. Forces such as gravity, friction, drag and angular drag also have effects on the movement of a vehicle. Compared to real world environment, time in Unity3D is discrete (default simulation time step is 0.02 s), so the accumulative force acts on the vehicle's rigid body at the start of each simulation time step, and resets to zero before the start of the next simulation time step.

Colliders are components defined by Unity3D to simulate physical collisions between two rigid bodies. Since colliders are the major physical parts of game objects, they also define the shapes and sizes of game objects in the simulation. Wheel colliders, specifically, are colliders of game objects that interact with the simulation environment in Unity3D. In our cooperative on-ramp merging case study, we adopt two different vehicle models with realistic dimensions in the real world, which are shown in FIGURE 2 (a) and (b).

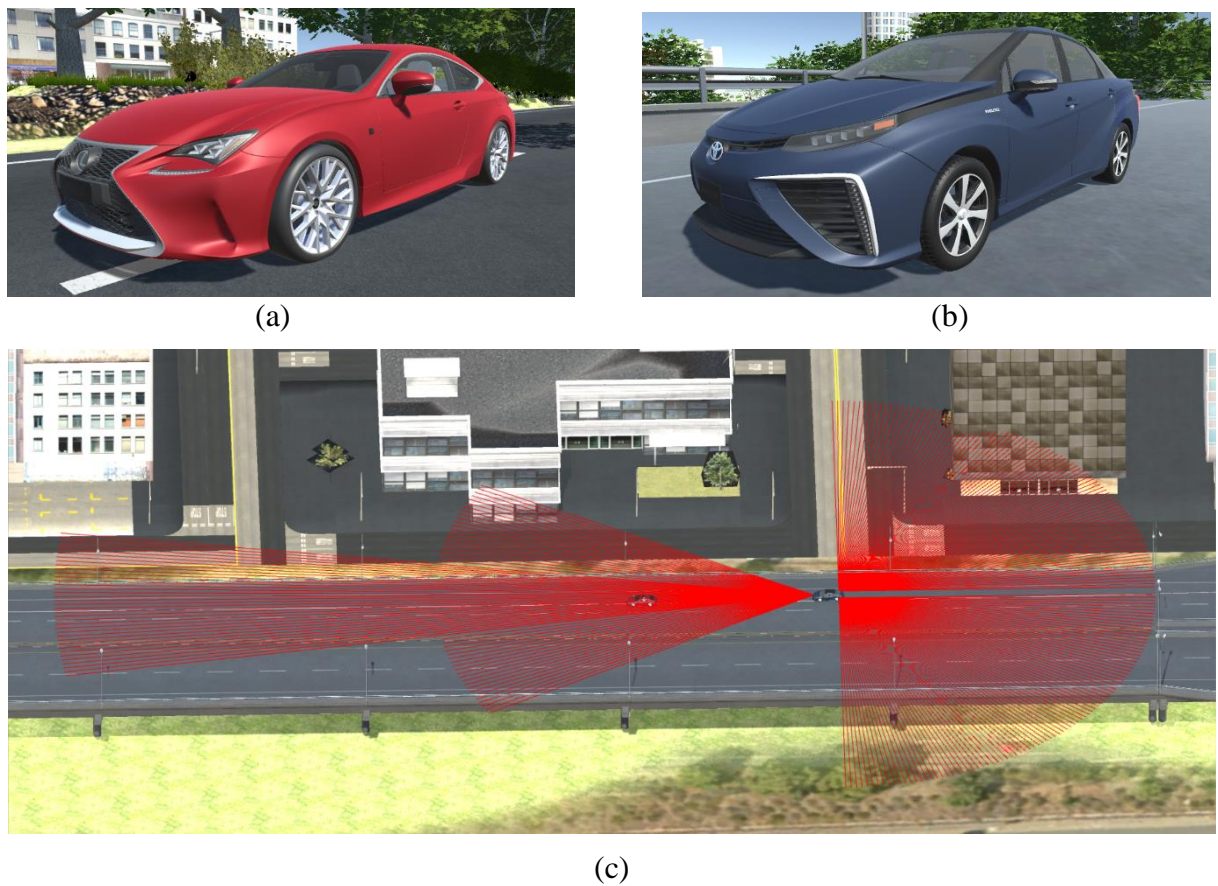

(a)　　　　　　　　　　　　　　　　(b)

(c)

**FIGURE 2 Vehicle Models and Radar Illustration Built in Unity3D.**

In order to enable vehicles with CAV technology in Unity3D, scripts with cooperation protocol are attached to enable their connectivity, and sensors are integrated to enable their autonomy. Specifically, a script written in Unity3D's C#-based Mono Scripting API is attached to all merging vehicles, which allows those CAVs to retrieve information from the infrastructure and other vehicles through V2X communications. This script also controls CAV's longitudinal movements by the proposed distributed consensus algorithm. Since we mainly focus on the longitudinal control of CAVs in this case study, the Simple Waypoint System is adopted, where a vehicle can track the preset trajectory with a user-defined longitudinal speed (*25*). Additionally, four radar sensors, including long/short-range front radars and left/right blind spot radars, are equipped on each CAV. As shown in FIGURE 2 (c), the long-range front radar has a relative narrow angle and long detection distance, so it is appropriate to be considered the primary front sensing approach. In order to prevent any rear-end collisions, the long-range front radar and the short-range front radar (redundant sensor) continuously check the distance to the physical preceding vehicle on the same lane.

**Simulation Environment Setup**
In order to conduct ABMS for CAVs, we build a simulation environment in Unity3D partially based on the city of Mountain View, California as shown in FIGURE 3 (a). California State Route 237 (CA-237) is the major corridor in this environment, with several on-ramps and off-ramps connecting it with urban arterial roads. In this work, we conduct the case study based on the on-ramp which connects E Middldfield Rd and CA-237 westbound. As shown in FIGURE 3 (b), this on-ramp has a length of 267 m, with a roadside unit-equipped infrastructure positioned between the on-ramp and highway. There is also an elevation difference between the on-ramp and highway, which means the vision of the on-ramp vehicle's driver is obstructed for a long period before merging.

In this simulation environment, a game object associated with a sphere collider is used to simulate the V2I-enabled infrastructure. Basically, the center of the sphere collider is positioned at this game object (which is represented by a power tower in FIGURE 3 (b)), and the sphere collider's radius can be set as the V2I communication range. The 'isTrigger" function of the sphere collider is enabled to prevent any collisions with incoming vehicles, otherwise vehicles cannot enter the volume of this collider. After setting this sphere collider to "isTrigger", when a vehicle enters and exits its volume, it sends "OnTriggerEnter" and "OnTriggerExit" messages. Then we can attach a configuration script to this game object to call these functions, and integrate the aforementioned vehicle sequencing protocol into the script. Namely, when a vehicle enters the radius of this sphere collider (i.e., enters the V2I communication range of the infrastructure), the "OnTriggerEnter" function is called, and the infrastructure retrieve information from the vehicle through "GetComponent" function. The infrastructure then processes this information along with other information retrieved from all other entering vehicles during a certain time window, and sends the sequence identification number back to each vehicle at the next time step of sorting process. Once a vehicle exits the radius of this sphere collider, the "OnTriggerExit" function is called, and the infrastructure clear all stored information of this vehicle.

The information sent from a vehicle to the infrastructure includes its longitudinal speed, acceleration, and the global position. Recall the proposed vehicle sequencing protocol, the distance to the merging point of a vehicle is needed to calculate its estimated arrival time. Therefore, we also come up with a map matching system to convert the global position of a vehicle into its distance to the merging point. Since the road segment is neither straight nor flat, we cannot simply calculate the distance between the vehicle's position and the merging point's position. Instead, we build paths with multiple waypoints along the lanes of both highway and on-ramp. Whenever the infrastructure gets the global position of a vehicle, it firstly compares the position with all waypoints' positions on that path to figure out which path segment this vehicle is currently on and what the next waypoint is. Once finished, the distance to the merging point of this vehicle is the sum of its distance to the next waypoint and the path length from the next waypoint to the merging point.

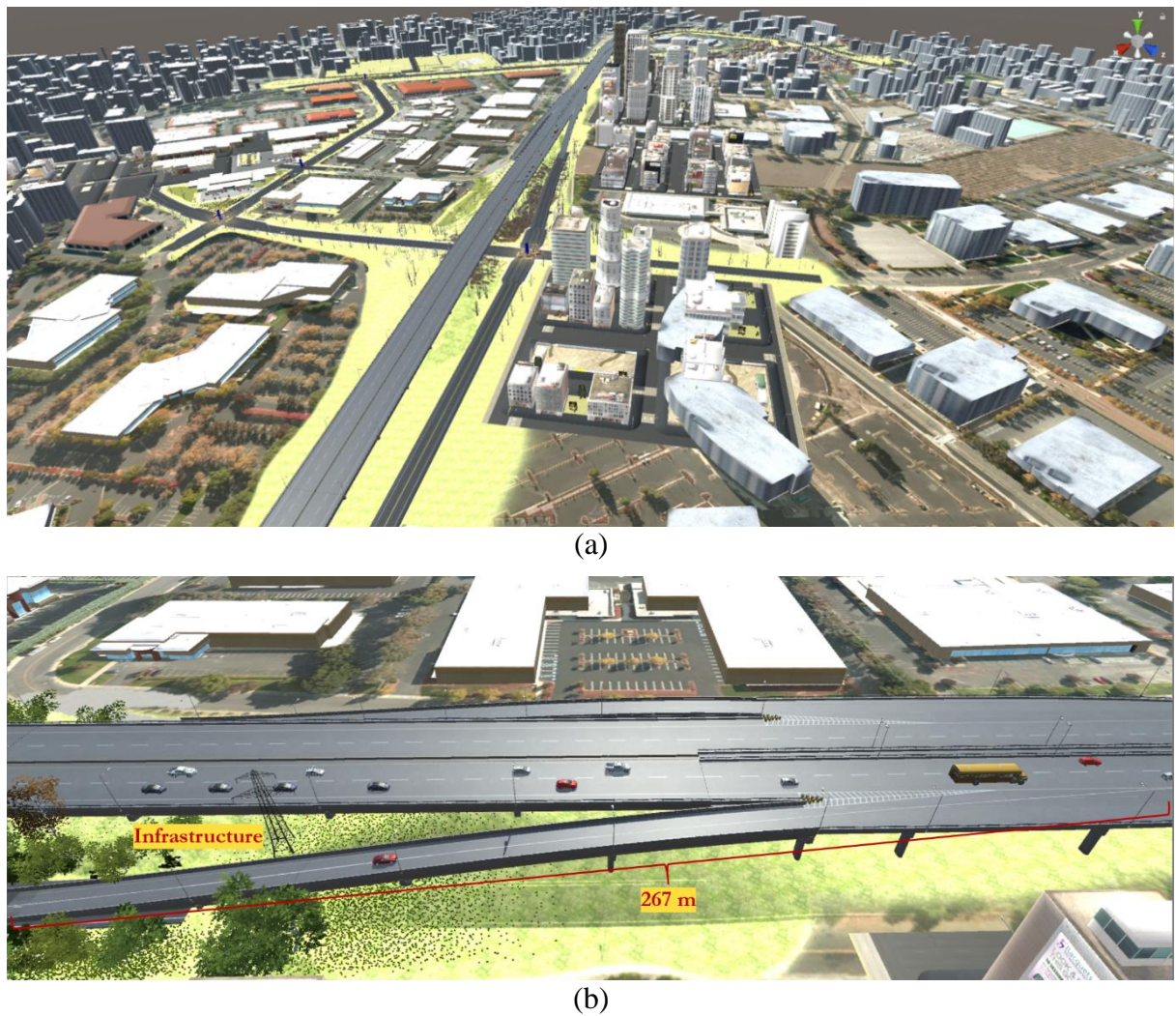

**FIGURE 3 Game Environment of Mountain View, California Built in Unity3D.**

It should be noted that building such a CAV simulation environment which conforms to various test criteria requires huge efforts from developers. Test criteria are oftentimes decided by different OEMs, CAV features to be tested, and the level of confidence obtained from previous tests. In order to construct similar realistic test environment in the virtual simulation in an automatic manner, one needs to come up with some test case generation protocol, which is discussed in our previous work (*9*).

**Simulations and Results**
In this work, we study the case where one on-ramp vehicle tries to merge with a six-vehicle string traveling on the highway. The proposed cooperative on-ramp merging protocol is applied to all these seven vehicles. All other vehicles in this simulation are running with a cruise speed on their own lanes, and will not affect the cooperative merging process, hence they are not considered when we analyze simulation results. The on-ramp vehicle is discharged from the starting point of the on-ramp at an initial speed of 5 m/s. The highway vehicles are discharged from the upstream of the highway (which is outside the V2I communication range of the infrastructure) with random initial speeds and longitudinal positions. Before they reach the V2I communication starting point, they are driven in a CACC mode at a desired speed of 30 m/s and with a desired time-gap of 0.5 s. Upon highway vehicles arriving at the V2I communication starting point, the CACC string has already been formed with a stable state.

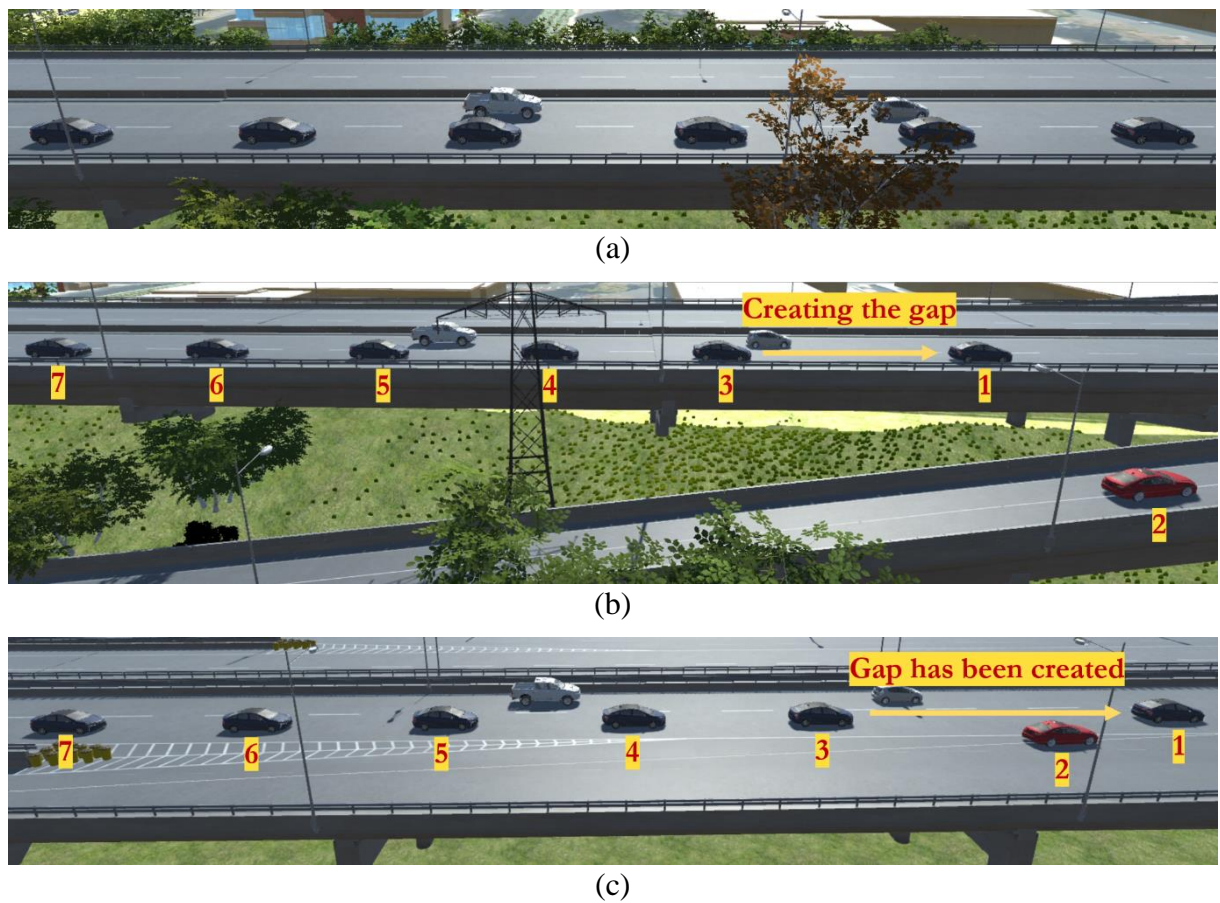

**FIGURE 4 Cooperative On-Ramp Merging Process.**

FIGURE 4 illustrates the cooperative on-ramp merging process in this simulation. FIGURE 4 (a) shows the stage when highway vehicles have already formed the CACC string, but have not entered the V2I communication range yet. FIGURE 4 (b) shows the on-ramp vehicle and all highway vehicles are already inside of the V2I communication range, and a sequence identification number of "2" is assigned to the on-ramp vehicle. Therefore, the highway vehicle with a sequence identification number of "3" is decelerating to follow the on-ramp vehicle, creating a gap for the on-ramp vehicle to merge. When the on-ramp vehicle is about to merge, as shown in FIGURE 4 (c), the gap has already been created and no further longitudinal speed adjustments are needed.

After running the simulation with all vehicles controlled by the proposed distributed consensus-based protocol, we also conduct human-in-the-loop simulations where the merging vehicle is controlled by a human driver on a driving simulator (FIGURE 5). In this scenario, the on-ramp vehicle is a conventional vehicle with no connectivity and autonomy, while all six highway vehicles are still CAVs. A highway vehicle can sense the on-ramp vehicle by its long-range front radar once the on-ramp vehicle cuts in front.

A car user control script and a car controller script work together to allow a human driver to control the longitudinal and lateral movements of the vehicle. The car user control script gets horizontal input from the driving simulator's steering wheel, and vertical input from the driving simulator's throttle/brake pedal. Then it calls the move function of the car controller script with the horizontal and vertical input. Wheel colliders of the vehicle will then be controlled based on these inputs, and vehicles can be set in motion. In order to reduce any system bias on the results of human-in-the-loop scenario, we recruit four subjects to drive the on-ramp vehicle, each for five times. The drivers drive the vehicle on the driving simulator based on their own preferences, namely, there is no requirement for them to drive aggressively or cautiously. We categorize their driving behaviors only after the speed trajectories are generated by their simulation runs.

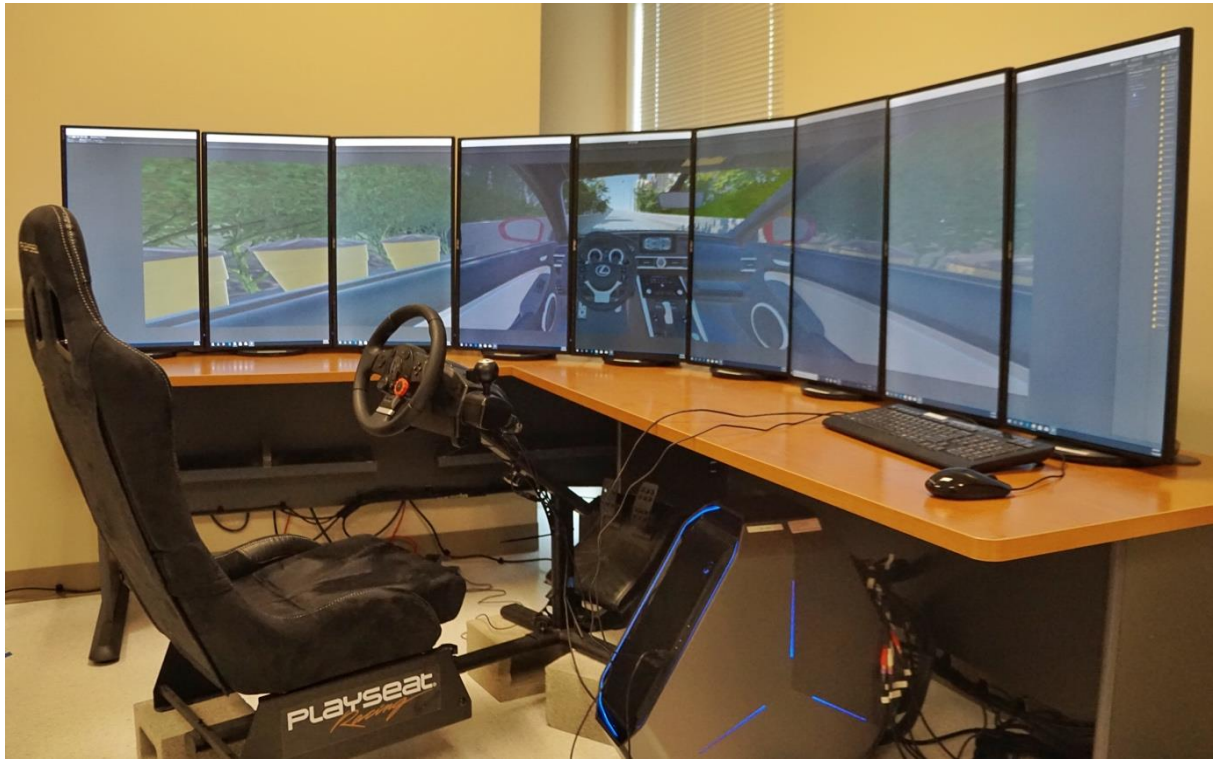

**FIGURE 5 Driving Simulator Platform.**

The results of vehicle speed profiles of the cooperative merging scenario and the human-in-the-loop scenario are compared in FIGURE 6. It should be noted that, we only show five out of twenty simulation results of the human-in-the-loop scenario in FIGURE 6, but all twenty simulation runs are considered when calculating results shown in TABLE 2. FIGURE 6 (a) shows the speed profiles generated from the proposed cooperative merging scenario, where highway vehicle 2 reaches the V2I communication starting point around 3 s, and it is assigned a sequence identification number of 3, which means it needs to follow the movement of the on-ramp vehicle. When vehicle 2 starts to decelerate to adjust its speed and longitudinal with respect to the on-ramp vehicle, its followers (highway vehicle 3, 4, 5, and 6) also decelerate accordingly since they are still in a CACC string. Meanwhile, the on-ramp vehicle gradually adjusts its speed and longitudinal position with respect to highway vehicle 1. Therefore, when the merging behavior happens around 18 s, there is no speed change of any vehicle at all.

For typical cautious driver scenario cases shown in FIGURE 6 (b) and (c), the on-ramp vehicle speeds up with a relatively small acceleration while it is on the on-ramp. Upon approaching the merging point, the on-ramp vehicle is already behind all highway vehicles, so it attaches to the end of the highway CACC string, and there is no speed change of any highway vehicle at all. For typical aggressive driver scenario cases shown as FIGURE 6 (d), (e) and (f), the on-ramp vehicle speeds up with a relatively large acceleration at first, since the driver's vision is obstructed and the driver does not know the traffic condition on the highway at that time. Once the driver observes the traffic condition around 14 s, the vehicle begins to slow down to adjust its speed with highway vehicles. Upon merging in the highway CACC string, all followers in that string also decelerate accordingly to prevent any rear-end collisions, and there are also some speed fluctuations afterwards.

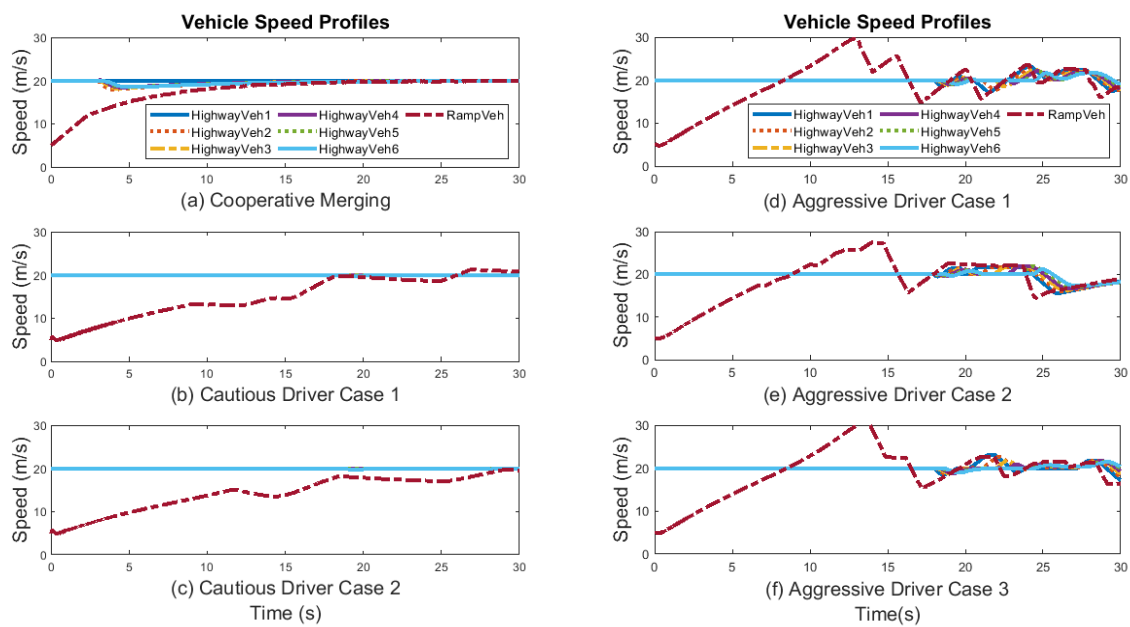

**FIGURE 6 Speed Profiles of Vehicles Driven in Different Scenarios.**

We also compare the system performance of cooperative merging scenario with the human-in-the-loop scenario, in terms of travel time, energy consumption and pollutant emissions. The number shown in each cell of TABLE 2 is the summation of all seven merging vehicles' results derived from the same traveling distance (600 m for each vehicle), and results of the human-in-the-loop scenario are calculated as average of all twenty simulation runs. Specifically, the travel time is a measurement to show the mobility benefit of the proposed cooperative on-ramp merging protocol, and it is calculated as the average time spent by all seven vehicles to travel through the same distance (600 m). It is shown in TABLE 2 that a reduction of 6.61 % can be generated by applying the protocol. Energy consumption and pollutant emissions are calculated by the U.S. Environmental Protection Agency's Motor Vehicle Emission Simulator (MOVES) (*26*). Compared to the human-in-the-loop scenario, the cooperative merging protocol introduces 7.82 % savings on energy consumption, and up to 58.37 % reduction on pollutant emissions, respectively.

**TABLE 2 Comparison results between cooperative merging and baseline**

|  | Travel Time | Energy | HC | CO | CO2 | NOx |
|---|---|---|---|---|---|---|
| **Cooperative Merging** | 218.14 s | 9153.97 KJ | 0.0094 g | 1.1737 g | 651.287 g | 0.0440 g |
| **Human-in-the-loop** | 233.58 s | 9930.56 KJ | 0.0200 g | 2.8192 g | 706.5392 g | 0.0759 g |
| **Reduction Percentage** | 6.61 % | 7.82 % | 53.00 % | 58.37 % | 7.82 % | 42.03 % |

**CONCLUSIONS AND FUTURE WORK**
In this work, we conducted ABMS of CAVs using the game engine Unity3D, due to Unity3D's visualization capability and many other advantages. Different from existing on-ramp merging methodologies, the distributed consensus-based cooperative on-ramp merging protocol was introduced. In the case study, agent-based models of CAVs were built in Unity3D with colliders and C#-based scripting API, and they were simulated in the Mountain View simulation environment. A comparison between the cooperative merging scenario with the human-in-the-loop scenario was made, analyzing the benefits of introducing the proposed protocol in terms of travel time, energy consumption and pollutant emissions.

Future work includes developing multi-fidelity CAV models with higher quality, allowing greater flexibility in the design process. Additional case studies of CAV technology can be conducted within this ABMS platform, such as CACC and eco-driving at signalized intersections. Integrating Unity3D with other software platforms via UDP socket communication model might lead to another research direction.

**ACKNOWLEDGEMENT**
The authors deeply appreciate Jonathan Shum's help on building the simulation environment in Unity3D. The authors also thank the participation of Yecheng Zhao, Muhammed Sayin and Chanwook Oh as drivers in the human-in-the-loop scenario, and the feedback from Dr. Baik Hoh to improve the quality of the paper.